\documentstyle [12pt]{article}
\def\be{\begin{equation}}
\def\ee{\end{equation}}
\def\bea{\begin{eqnarray}}
\def\eea{\end{eqnarray}}
\def\pd{\partial}

\begin{document}

\begin{titlepage}

\title{The General Solution of the Complex Monge-Amp\`ere  Equation
         in two dimensional space}

\author{D.B. Fairlie\\
{\it Department of Mathematical Sciences}\\
{\it University of Durham, Durham DH1 3LE}\\
and A.N. Leznov\\
{\it  Institute for High Energy Physics, 142284 Protvino,}\\{\it Moscow Region,
Russia}
{\it and}\\ 
{\it  Bogoliubov Laboratory of Theoretical Physics, JINR,}\\
{\it 141980 Dubna, Moscow Region, Russia}}
\maketitle

\begin{abstract}

The general solution to the Complex Monge-Amp\`ere equation in a two dimensional
space is constructed.

\end{abstract}

\end{titlepage}

\section{Introduction}

 The Complex Monge-Amp\`ere equation in $2$-dimensional space takes the form:
\bea 
\det\left|\begin{array}{cc}
\frac{\pd^2\phi}{\pd y_1\pd\bar y_1} & \frac{\pd^2\phi}{\pd y_1\pd\bar y_2}\\
\frac{\pd^2\phi}{\pd y_2\pd\bar y_1} & \frac{\pd^2\phi}{\pd y_2\pd\bar y_2}\\
\end{array}\right|\,=\,0.\label{batman}
\eea
Its real form, which arises from (\ref{batman}) under the assumption that the solution
depends only upon $2$ arguments $x_i=y_i+\bar y_i$ was solved before by different 
methods \cite{renat},\cite{dbf}. But to the best of our knowledge the general solution
of the complex M-A equation (\ref{batman}) is not known. 

In the present paper we follow two different aims. On the one hand we want to
fill the gap and using the method of our paper \cite{dbf} (the solution of the 
real M-A 
equation in a space of arbitrary dimension) obtain and present the general  
solution of the complex version of M-A equation (\ref{batman}) in  implicit 
form. On the other hand we want to prepare the reader to understand the 
methods for obtaining the general solution of the homogeneous M-A equation in a
space of the arbitrary dimension. This work is now nearly finished.

\section{Equivalent First Order Equations}

The complex M-A equation, (\ref{batman}) is the eliminant of $2+2$ linear
equations (from the linear dependence between rows or columns of the determinant matrix),
which may be written as:
\be \sum^2_{i=1} \alpha^i \phi_{y_i,\bar y_k}=0,\quad 
\sum^2_{i=1} \beta^i \phi_{\bar y_i,y_k}=0\label{1}
\ee
where $\phi_{y_i}$ denotes $\displaystyle{\frac{\pd\phi}{\pd y_i}}$ etc.

The next  rows contain an obvious transformation (\ref{1}) using
only  the rules of differentiation and some new definitions:
\be
(\sum_{i=1}^2 \alpha^i \phi_{y_i})_{\bar y_k}=
\sum_{i=1}^2 \alpha^i_{\bar y_k} \phi_{y_i}\quad
(\sum_{i=1}^2 \beta^i \phi_{\bar y_i})_{y_k}=
\sum_{i=1}^2 \beta^i_{y_k} \phi_{\bar y_i}\label{R}
\ee
Introducing two new functions
$$
R=\sum_{i=1}^2 \alpha^i \phi_{y_i},\quad 
\bar R=\sum_{i=1}^2 \beta^i \phi_{\bar y_i}
$$
and considering them as a functions of three arguments $R=R(\alpha,y),
\bar R=\bar R(\beta,\bar y)$ (under the assumtion that $\det J(\alpha,\bar y)$ 
and $\det J(\beta,y)$ are different from zero), we rewrite equations (\ref{R}) 
in an equivalent form:
\be
R_{\alpha_i}=\phi_{y_i},\quad \bar R_{\beta_i}=\phi_{\bar y_i}\label{RR}.
\ee
Multiplying each equation (\ref{RR}) respectively by $\alpha_i,\beta_i$,
summing the results (recalling the definitions of the functions
$R$,$\bar R$, we come to the conclusion that they (with respect to $\alpha,\beta$ arguments)
are homogeneous functions of degree 1. Introducing the notation
$$
\frac{\alpha_1}{ \alpha_2}=u,\quad \frac{\beta_1}{ \beta_2}=v,
$$ 
we can represent the dependence of the functions $R,\ \bar R$  in the following 
way:
$$
R=\alpha_2 R(u;y_1,y_2),\quad \bar R=\beta_2 \bar R(v;\bar y_1,\bar y_2) 
$$
Substituting these expressions into equations (\ref{RR}) we arrive at the
following relations which form the backbone of our further investigations
\be
\phi_{y_1}=R_u,\quad \phi_{y_2}=R-uR_u\label{ME1}
\ee
\be
\phi_{\bar y_1}=\bar R_v,\quad \phi_{\bar y_2}=\bar R-v \bar R_v
\label{ME2}
\ee

\section{Conditions of selfconsistency (part I)}

Using the condition of the equivalence of second mixed partial derivatives taken in 
different orders, we shall be able to separate the main system (\ref{ME1}),
(\ref{ME2}) and extract from it a very important system of equations connecting
the functions $u,v$ only. To this end let us calculate and equate the second
mixed derivatives of the following pairs of variables $(\bar y_1,y_1),
(\bar y_1,y_2),(\bar y_2,y_1),(y_2,\bar y_2)$.

We have in consequence for the pair $(\bar y_1,y_1)$:
$$
(R_u)_{\bar y_1}=(\bar R_v)_{y_1},\quad R_{u,u}u_{\bar y_1}=\bar R_{v,v}v_{y_1};
$$
for the pair $(y_1,\bar y_2)$:
$$
(R_u)_{\bar y_2}=(\bar R-\bar R_v\bar R_v)_{y_1},\quad R_{u,u}u_{\bar y_2}=-  
\sum v\bar R_{v,v}v_{y_1};
$$
for the pair $(\bar y_1 ,y_2)$:
$$
\bar R_{v,v}v_{y_2}=- u R_{u,u}u_{\bar y_1};
$$
and finally the pair $(y_2,\bar y_2)$, leads to equations
$$
uR_{u,u}u_{\bar y_2}=v\bar R_{v,v}v_{y_2}
$$
Multiplying first equations respectively by $u,v$, summing
the results and comparing with the second and the third sets of equations 
respectively we come to the following separate system of equations (assuming that $R_{u,u}\neq 0,
\bar R_{v,v}\neq 0)$,  which the functions 
$(u,v)$satisfy:
\be
u_{\bar y_2}+vu_{\bar y_1}=0,\quad  
v_{y_2}+uv_{y_1}=0\label{FUV}
\ee
The system (\ref{FUV}) was solved before \cite{dbf},\cite{rev} but for the convenience 
of the reader the next two sections will be devoted to its consideration.

The last comment is the following; the hydrodynamic type  system (\ref{FUV})
is the result of only $2$ equations of second mixed derivatives. Namely it arises from
combinations of the first, second and third. It is not dificult to
check that the equation for the pair $(y_2,\bar y_2)$ automatically satisfies (\ref{FUV}) also.
Thus only one equation  remains unsolved, that connecting the pairs with barred and 
unbarred index 1. The consequences  will be considered in section 6.

\section{System of hydrodynamic type }

We understand by a system of hydrodynamic type the system of equations 
(\ref{FUV})  
rewritten below:
\be
v_{y_2}+u v_{y_1}=0,\quad 
u_{\bar y_2}+v u_{\bar y_2}=0\label{I}
\ee

Two properties of this system will be crucially important in what
follows.

Proposition 1.

The pair of operators:
\be
D=\frac{\partial}{\partial y_2}+u \frac{\partial}{\partial y_1},
\quad 
\bar D=\frac{\partial}{\partial \bar y_2}+v \frac{\partial}{\partial \bar y_1}
\label{3}
\ee
are mutually commutative if $(u,v)$ are solutions of the system (\ref{I}).

Acting with the help of operators $(D,\bar D)$ on the second and the first 
equations of (\ref{I}) respectively we come to the conclusion that the two
functions:
\be
\bar D(v)=v{\bar y_2}+v v{\bar y_1},\quad 
D(u)=u_{y_2}+u u_{y_1}
\ee
are also solutions of the first and the second system of equations (\ref{I}).

As a corollary we obtain the following; 

Proposition 2.
\be
v_{\bar y_2}+v v_{\bar y_1}=V(v;\bar y_1,\bar y_2),\quad 
u_{y_2}+u u_{y_1}=U(u;y_1,y_2)\label{BA}
\ee
Indeed the $2$ sets of variables $(1,u)$, and $(1,v)$ respectively satisfy  a 
linear system of algebraic equations of $2$ terms, the matrix of which 
coincides with the Jacobian matrix 
\be
J=\det_2 \left|\begin{array}{cc} 
v & V \\
y_1 & y_2 \end{array}\right|\nonumber
\ee
which in the case of a non-zero solution of the linear system must
vanish. So Proposition 2 is proved.

In comparison  with (\ref{I}), (\ref{BA}) is  an inhomogeneous system of hydrodynamic
equations separated into functions  $(u,v)$.
\section{General solution of the hydrodynamic system}

Suppose we have the equation defining implicitly unknown 
function $(\psi)$ in $(4)$ dimensional space $(y,\bar y)$:
\be
Q(\psi;y)=P(\psi;\bar y)\label{D}
\ee
where $Q,P$ are arbitrary functions of its 3 arguments.

With the help of the usual rules of differentiation of implicit functions
we find from (\ref{D}):
\be
\psi_y=(P_{\psi}-Q_{\psi})^{-1} Q_y,\quad \psi_{\bar y}=-(P_{\psi}-Q_{\psi})^
{-1}P_{\bar y}\label{DD}
\ee
Let us assume, that between two derivatives with respect to the barred and 
unbarred variables there exists the following linear dependence:
\be
\sum^2_1 c_i \psi^{\alpha}_{y_i}=0,\quad \sum^2_1 d_i \psi^{\alpha}_{\bar y_i}=0
\label{LC}
\ee
and analyse the consequences following from these facts.

Assuming that $c_2\neq 0,d_2\neq 0$, dividing  them into each 
equation of the left and right systems  respectively and introducing the 
notation $u={c_1\over c_2},v={d_1\over d_2}$ we rewrite the last 
systems in the form:
\be
\psi_{y_2}+u \psi_{y_1}=0,\quad \psi_{\bar y_2}+v\psi_{\bar y_1}=0\label{MS} 
\ee
Substituting the values of the derivatives from (\ref{DD}) and multiplying  the
result by $(P_{\psi}-Q_{\psi})$ we obtain:
\be
Q_{y_2}+u Q_{y_1}=0,\quad P_{\bar y_2}+vP_{\bar y_1}=0\label{D1}
\ee

From the last equations it immediately follows that:
\be
u=-{Q_{y_2}\over Q_{y_1}},\quad v=-{P_{\bar y_2}\over P_{\bar y_1}} \label{UV}
\ee
We see that if we augment the initial system (\ref{D}), by two functions
$(u,v)$ defined by (\ref{UV}) then the operators of differentiation $D,\bar D$
defined by (\ref{3}) in connection with (\ref{MS})  annihilate each 
$\psi$ either as $Q,\ P$ functions:
\be
D \psi=\bar D \psi=D Q=D P=\bar D Q=\bar D P=0 \label{VIC}
\ee
This means that $D\bar f(\psi,\bar y)=\bar D f(\psi, y)=0$. And as a direct
corollary of this fact $Dv=\bar D u=0$ and so the generators $D,\bar D$ which 
have been constructed commute.

Thus we have found the general solution of the hydrodynamic
system and a concrete realisation of the manifold with the properties of the 
previous section.

With respect to generators $D,\bar D$ all functions of $4$ dimensional space
$(y_i,\bar y_k)$ may be divided into the following subclasses: functions of  
general position $F, D F\neq 0,\bar D F\neq 0$, the holomorphic functions 
$f,\bar D f=0, D f\neq 0$, antiholomorphic ones $\bar f, D \bar f=0,\bar D 
\bar f\neq 0$ and $f^0$ "central" holomorphic and antiholomorphic 
simultaneously; $\bar D f^0=D f^0=0$. 
 
Each central function may be represented in 
the form:
$$
f^0=f^0(Q)=f^0(P)=g^0(\psi)
$$

\section{Conditions of selfconsistency (part II)}
Here we compare and equate second mixed derivatives of the pairs $(y_1,y_2),
(\bar y_1,\bar y_2)$ and remaining from the section part I the pair 
$(y_1\bar y_1)$. All calculations are straightforward.

The conditions of selfconsistency of the pairs $(y_1,y_2),(\bar y_1,\bar y_2)$ 
may be manipulated into the following compact form:
\be
D R_u=R_{y_1},\quad \bar D \bar R_v=\bar R_{\bar y_1}\label{F2}
\ee

There is only one equation of selfconsistency, connecting the barred and 
unbarred index $1$:
\be
(R_u)_{\bar y_1}=(\bar R_v)_{y_1},\quad R_{uu}u_{\bar y_1}=\bar R_{vv}v_{y_1},
\quad R_{uu}u_{\psi}\psi_{\bar y_1}=\bar R_{vv}v_{\psi}\psi_{y_1}\label{F4}.
\ee

\section{Solution of selfconsistency equations}
Substituting into (\ref{F4}) the known values of the derivatives of
$\psi$ function (\ref{DD}) we pass to the final equation of interest:
\be
{R_{uu}u_{\psi}\over Q_{y_1}}=-{\bar R_{vv}v_{\psi}\over P_{\bar y_1}}=A^0
_{\psi}\nonumber
\ee
Indeed the left hand side of the last equality is a holomorphic function, 
the right hand side an antiholomorphic one. Thus $A^0_{\psi}$ is a central 
function.

Considering now $R_u=R_u(\psi;y_1,y_2)$ and $\bar R_v=\bar R_v(\psi;\bar y_1,
\bar y_2)$ we solve the equations containing the function $A^0_{\psi}$  in 
the form:
\bea
R_u&=&\Theta_{y_1}(A;y_1,y_2),\quad Q=\Theta_A(A;y_1,y_2)\nonumber\\
R_v&=&\bar \Theta_{\bar y_1}(A;\bar y_1,\bar y_2),\quad P=-\bar \Theta_A(A;
\bar y_1,\bar y_2)\label{RT}
\eea

It remains only to check equalities (\ref{F2}). Let us distinguish by
upper indices $u,A$ the  corresponding 
derivatives
$\frac{\partial^u}{\partial y_i},\frac{\partial^A}{\partial y_i}$  on the
 space coordinates $(y,\bar y)$ taken keeping
$u,(v)$ constant in first case and the function $A$ constant in the second. The
 equality 
which has to be checked in this notation has the form:
\be
\frac{\partial^u}{\partial y_1} R_u=\frac{\partial}{\partial u}
\frac{\partial^u}{\partial y_1} R\nonumber
\ee
Keeping in mind that $D^u A=0$ ($A$ is a central function) and the definition of
all values involved in terms of the function $\Theta$  we obtain in consequence for 
the right hand side of the last equality
$$
(D^u R_u)_u=(D^u \frac{\partial^A}{\partial y_1}\Theta)_u=
(D^u \frac{\partial^A}{\partial y_1}\Theta)_A A_u=
$$
$$
\left((\frac{\partial^{2A}}{\partial y_1\partial y_2}+u\frac{\partial^{2A}}
{\partial y_1\partial y_1})\Theta\right)_A A_u=\Theta_{y_1,y_1}+(\Theta_{A,y_1,y_2}-
{\Theta_{A,y_2}\over \Theta_{A,y_1}}\Theta_{A,y_1,y_1}) A_u=
$$
$$
\Theta_{y_1,y_1}+\Theta_{A,y_1} A_u \frac{\partial^Au}{\partial y_1}
$$
In all transformations above we have not written  the  upper index $A$ with respect
to derivatives of the space coordinates $y$.

Similar calculations for the left hand side leads to:
$$
\frac{\partial^u}{\partial y_1} \Theta_{y_1}=\Theta_{y_1,y_1}+\Theta_{A,y_1} A_u
\frac{\partial^Au}{\partial y_1}  
$$
which shows that equalities (\ref{RT}) are satisfied.
But (\ref{F2}) in its turn is an equation of second order with respect to the 
unknown  function $R$. We rewrite it in explicit form substituting instead of 
$R_u$ its value from (\ref{RT}):
$$
(\frac{\partial^{2}}{\partial y_1\partial y_2}+u\frac{\partial^{2}}
{\partial y_1\partial y_1})\Theta=R^u_{y_1}\equiv R_{y_1}+R_A A_{y_1}=
$$
$$
R_{y_1}-{u_{y_1}\over u_A}R_A=R_{y_1}-u_{y_1}R_u=R_{y_1}-u_{y_1}\Theta_{y_1}
$$
In the process of evaluation of the last expression the crucial step was the
calculation of $A_{y_1}$  keeping  $u$ at a fixed value. It was achieved by 
direct differentiation of the definition of $u$ rewritten in the form:
$$
\Theta_{A,y_2}+u\Theta_{A,y_1}=0
$$
with respect to the argument $y_1$  (with fixed $u$) and regroupping the
terms arising.

Preserving in the last equality the first and the last terms we obtain the
equations for the function $R$  in  integrable form. The result of  integration
determines the function $R$ in terms of the function $\Theta$  in a very 
attractive form:
$$
R=D \Theta,\quad \bar R=\bar D \bar \Theta
$$
Substituting these expression in equations connecting derivatives of the solution 
of the M-A equation (\ref{ME1}), (\ref{ME2}) with the
functions $R,\bar R$  we obtain 
finally:
$$ 
\phi_{y_1}=\Theta_{y_1},\quad \phi_{y_2}=\Theta_{y_2},\quad \phi_{\bar y_1}=
\bar \Theta_{\bar y_1},\quad \phi_{\bar y_2}=\bar \Theta_{\bar y_2}
$$

\section{Concluding remarks}
The main result of the present paper is in the theorem of the previous section.
But not less important is the hydrodynamic like system of equations (\ref{UV})
solved in the middle  of the calculations. It effectively defines the manifold of 
solutions of the complex M-A equation in two dimensions. In the final result
it is encoded in the equation determining the function $A$. In the one 
dimensional limit
it does not pass directly to the Monge equation but another new integrable
system:
$$
u_t+vu_x=0,\quad v_t+uv_x=0
$$ 
The Monge equation is contained among  special cases of this system under the reduction $u=v$.

In the near future using the scheme of the present paper we are going to publish
a similar result in connection with the M-A equation in a space of arbitrary 
dimension.

\section*{Acknowledgements}

One of the authors (ANL) is indebted to the Center for Research on Engineering
and Applied Sciences (UAEM, Morelos, Mexico) for its hospitality and
Russian Foundation of Fundamental Researches (RFFI) GRANT N 98-01-00330 for 
partial support.


\begin{thebibliography}{3}
\bibitem{renat} R.Zhdanov,  On the general solution of the multi-dimensional
Monge-Amp\'er\'e  equation, in: {\em Symmetry Analysis and Solutions of 
Mathematical Physics Equations}, Kyiv: Institute of Mathematics, 1988,
13-16.
\bibitem{dbf}D.B.Fairlie and  A.N. Leznov, General solutions of the Monge-Amp\`{e}re equation in  \hfill\break  $n$-dimensional space {\it Journal of Geometry and Physics}.
{\bf 16} (1995) 385-390
\bibitem{rev} D.B.Fairlie Integrable Systems in Higher Dimensions 
{\it Quantum Field Theory, Integrable Models and Beyond} 
Editors. T. Inami and R. Sasaki {\it Progress of Theoretical Physics Supplement}
 {\bf 118} (1995) 309-327.
\end{thebibliography}
\end{document}